%% file: main.tex
\documentclass{article}

\usepackage{microtype}
\usepackage{graphicx}
\usepackage{subcaption}
\usepackage{booktabs} 
\usepackage{xspace}
\usepackage{amssymb}
\usepackage{xcolor}
\usepackage{amsmath}
\usepackage{footnote}

\usepackage{hyperref}

\newcommand{\oursys}{AI Metropolis\xspace}

\newcommand{\GA}{GenAgent\xspace}

\usepackage[accepted]{mlsys2024}

\mlsystitlerunning{AI Metropolis}

\begin{document}

\twocolumn[
\mlsystitle{AI Metropolis: Scaling Large Language Model-based Multi-Agent Simulation with Out-of-order Execution}



\mlsyssetsymbol{equal}{*}

\begin{mlsysauthorlist}
\mlsysauthor{Zhiqiang Xie}{stf}
\mlsysauthor{Hao Kang}{gatech}
\mlsysauthor{Ying Sheng}{stf}
\mlsysauthor{Tushar Krishna}{gatech}
\mlsysauthor{Kayvon Fatahalian}{stf}
\mlsysauthor{Christos Kozyrakis
}{stf}
\end{mlsysauthorlist}

\mlsysaffiliation{stf}{Stanford University}
\mlsysaffiliation{gatech}{Georgia Institute of Technology}

\mlsyscorrespondingauthor{Zhiqiang Xie}{xiezhq@cs.stanford.edu}

\mlsyskeywords{Machine Learning, MLSys}

\vskip 0.3in

\input{0_abstract}

]



\printAffiliationsAndNotice{}

\input{1_introduction}

\input{2_background}

\input{4_design}

\input{5_evaluation}

\input{6_related}

\input{7_future}

\nocite{langley00}

\bibliography{reference}
\bibliographystyle{mlsys2024}

\appendix

\input{appendix}


\end{document}

%% file: 0_abstract.tex
\begin{abstract}
With more advanced natural language understanding and reasoning capabilities, large language model (LLM)-powered agents are increasingly developed in simulated environments to perform complex tasks, interact with other agents, and exhibit emergent behaviors relevant to social science research and innovative gameplay development.
However, current multi-agent simulations
frequently suffer from inefficiencies due to the limited parallelism caused by false dependencies, resulting in performance bottleneck.
In this paper, we introduce AI Metropolis, a simulation engine that improves the efficiency of LLM agent simulations by incorporating out-of-order execution scheduling. 
By dynamically tracking real dependencies between agents, \oursys minimizes false dependencies, enhancing parallelism and enabling efficient hardware utilization.
Our evaluations demonstrate that AI Metropolis achieves 
speedups from $1.3\times$ to $4.15\times$
over standard parallel simulation with global synchronization, approaching optimal performance as the number of agents increases.
\end{abstract}

%% file: 1_introduction.tex
\section{Introduction}

Large Language Models (LLMs) are advanced machine learning models trained on vast amounts of data, excelling in understanding and generating natural language. They have transformed natural language processing, enabling high-accuracy applications like text completion~\cite{merity2016pointer}, summarization~\cite{narayan2018dont}, and reasoning~\cite{DBLP:journals/corr/abs-2110-14168}. Beyond simple queries and chatbot interactions~\cite{chatgpt}, there is growing interest in using LLMs to create self-planning, decision-making, problem-solving, and reasoning engines~\cite{wang2024survey}. These advancements aim to develop human-like agents~\cite{xi2023rise} capable of performing complex tasks, interacting with environments and other agents, and making informed decisions based on context.

This interest is particularly pronounced in developing LLM-powered agents within simulated environments, where two unique opportunities arise.
First, simulation environments provide an efficient platform for testing and tuning LLM agents~\cite{dubois2024alpacafarm, liu2023training, wang2023describe}, with potential applications extending to real-world settings or virtual environments like gaming.
Second, the enhanced natural language understanding and reasoning capabilities of LLMs have sparked a trend of examining emergent social behaviors of these agents in game-like simulations~\cite{Park2023GenerativeAgents, altera_ai_npcs}. Such studies can serve as predictive models, forecasting real-world human behaviors, which is highly valuable for social science research~\cite{ziems2023can, grossmann2023ai}.

Despite the significance of simulation environments for LLM agents, the efficiency of managing simulation states and scheduling LLM requests in simulations are often overlooked, leading to slow and inefficient simulation processes.
Recent research work commonly implements their LLM agents simulation~\cite{Park2023GenerativeAgents, gong2023mindagent} directly adhering to a paradigm borrowed from reinforcement learning agent training and traditional multi-agent simulation~\cite{emau2011multi}, where simulation time is discretized into time steps and a \textit{step} (or similar) function is invoked to apply agents' actions, synchronize the environment, and coordinate agents at each interval.
This pattern, illustrated in Algorithm~\ref{alg:procedure}, is prevalent in prominent reinforcement learning frameworks such as OpenAI Gym~\cite{gym}, Meta Pearl~\cite{zhu2024pearl}, and TensorFlow Agents~\cite{TFAgents}.
The rationale behind this design is that global synchronization, enforced through the \textit{step} function, easily maintains \textit{temporal causality} within the simulation by serializing tasks along the simulation time axis. 

While this design suits the needs of reinforcement learning and traditional multi-agent simulations, we found it inefficient for LLM agents due to their unique performance characteristics, necessitating a new scheduling approach.
Simulations involving LLM agents, like other LLM-powered applications, are heavily dominated by inference time. 
Taking the pioneering work on generative agents~\cite{generativecode} (\GA) as an example, our trace analysis reveals that approximately $95\%$ of the simulation time is dedicated to LLM inference. 
Consequently, inference throughput becomes crucial, as higher throughput directly translates to shorter completion times and lower costs.
Furthermore, recent studies on LLM serving engines~\cite{kwon2023efficient, zheng2023efficiently} indicate that large batch sizes are essential for achieving high inference throughput.
Unfortunately, the traditional approach, which enforces step-wise temporal causality across simulation steps, introduces excessive synchronization that significantly reduces parallelism, thereby reducing achievable batch sizes and leading to low throughput.
This reduction in parallelism occurs because the execution times of LLM queries from different agents within a simulation step can vary significantly due to two main reasons: 
(1) variations in the input and output lengths of queries, and (2) differing numbers of queries sent by different agents.
As a result, enforcing global synchronization causes many agents to wait unnecessarily for each step to complete, limiting concurrent LLM queries and further reducing throughput.
This reduction in parallelism also hampers scalability, as adding more resources fails to meaningfully decrease the overall simulation completion time.

Our key observation is that temporal causality in simulations can be maintained without the costly global step-wise synchronization in the simulation. Intuitively, if two agents are far apart in a simulated world, the actions of one agent will not be immediately visible to the other. 
This means that it is often unnecessary for all agents to wait for each step to finish before proceeding, revealing a false dependency that can be removed to improve efficiency in the simulation.

To address the aforementioned challenge, in this paper, we present \oursys, a multi-agent simulation engine for LLM-powered agents that introduces the concept of out-of-order execution to simulation scheduling.
By carefully tracking real dependencies between agents during runtime, we can effectively eliminate most false dependencies. This approach allows certain agents to advance in simulation time ahead of others without affecting the simulation's outcome, which significantly enhances parallelism and thus better utilizes hardware with larger inference batch sizes.
Dependency tracking is achieved by analyzing the temporal-spatial relationships between agents, where the number of steps an agent can advance is determined by its distance from other agents. Similar to the scoreboard in out-of-order execution algorithms, \oursys maintains a specialized dependency graph to efficiently track these relationships.

\oursys provides LLM agent developers with interfaces similar to OpenAI Gym~\cite{gym}, while seamlessly managing simulation state updates, database I/O, scheduling, and LLM inference processes. 
We evaluated \oursys by replaying traces collected from instrumenting the original \GA implementation~\cite{generativecode} across different models, GPUs, and simulation scales.
The results demonstrate that \oursys outperforms the standard approach of parallel simulation with global step synchronization, 
achieving
speedups from $1.3\times$ to $4.15\times$, and approaching an order of magnitude improvement over the original \GA implementation.
As the number of agents increases, \oursys rapidly nears optimal performance, demonstrating its scalability and effective dependency management.
We plan to open-source \oursys to accelerate research in large-scale LLM agent simulation and release the collected traces to fill a critical gap in LLM serving benchmarks, particularly given the unique and complex dependency patterns among LLM calls.

\renewcommand*\footnoterule{}
\renewcommand{\thefootnote}{\arabic{footnote}}

\begin{minipage}{0.45\textwidth}
\renewcommand*\footnoterule{}
\renewcommand{\thefootnote}{\arabic{footnote}}

\begin{savenotes}
\begin{algorithm}[H]
    \caption{Traditional Simulation Scheduling 
    }\label{alg:procedure}

    \label{alg:algorithm}

    \begin{algorithmic}[1]
\STATE \textbf{Input:} target\_step, agents, world
\STATE \textbf{Initialize:} step $\gets$ 0
\WHILE{step $<$ target\_step}
    \STATE actions $\gets$ \texttt{[ ]}
    \FORALL{agent \textbf{in} agents}
        \STATE actions.\texttt{append}(agent.\textbf{\textit{proceed}}(world))
        \footnote{The \textit{\textbf{proceed}} function involves LLM calls to process tasks. 
        }
    \ENDFOR
    \STATE world.\textbf{\textit{step}}(actions)
    \STATE step $\gets$ step + 1
\ENDWHILE
\end{algorithmic}
\end{algorithm}
\end{savenotes}
\end{minipage}

%% file: 2_background.tex
\section{Simulation of LLM Agent Interaction}

\subsection{Background}
While simulation environments can be diverse with complex action spaces and interactions, the high-level procedure of a simulation generally follows a common pattern as outlined in Algorithm~\ref{alg:procedure}. In this pattern, agents determine future actions based on the current world states as well as their internal states, and these actions affect the world as the simulation progresses. Note the \textit{agent.proceed} and \textit{world.step} functions are defined by the agent developers and environment respectively, and can be implemented by manual rules or via invoking LLMs (or both) for decision making.

To illustrate the challenge of achieving efficient simulation, we use \GA as a concrete example within the broad family of simulations for LLM agent interaction. \GA proposes a comprehensive agent architecture and interaction mechanisms that have inspired many subsequent works. Their concepts and workflows are widely adopted in the community.
In \GA, 25 agents inhabit a world called SmallVille, akin to a grid-based game. Each agent possesses its own personality, social relationships, and daily routines. They navigate the world, interact with objects, and converse with other agents. 
The \textit{agent.proceed} function on line 6 of Algorithm~\ref{alg:procedure} is expanded to Algorithm~\ref{alg:proceed}
to detail several steps: perceiving surroundings, planning actions based on recent events, recalling relevant past events from memory, following a structured daily routine, and occasionally reflecting on their actions or experiences. Each of these steps can involve LLM calls.
Each step corresponding to ten seconds in simulated time in \GA.

\begin{algorithm}[H]
    \caption{Proceed function in \GA}\label{alg:proceed}

    \label{alg:algorithm}

    \begin{algorithmic}[1]
\STATE \textbf{Input:} agent, world
\STATE \textbf{Output:} action
\STATE perceived\_events $\gets$ agent.\textbf{\textit{perceive}}(world)
\STATE retrieved\_events $\gets$ agent.\textbf{\textit{retrieve}}(perceived\_events)
\STATE action $\gets$ agent.\textbf{\textit{plan}}(world, retrieved\_events)
\end{algorithmic}
\end{algorithm}

\begin{figure}
    \centering
    \includegraphics[width=\linewidth]{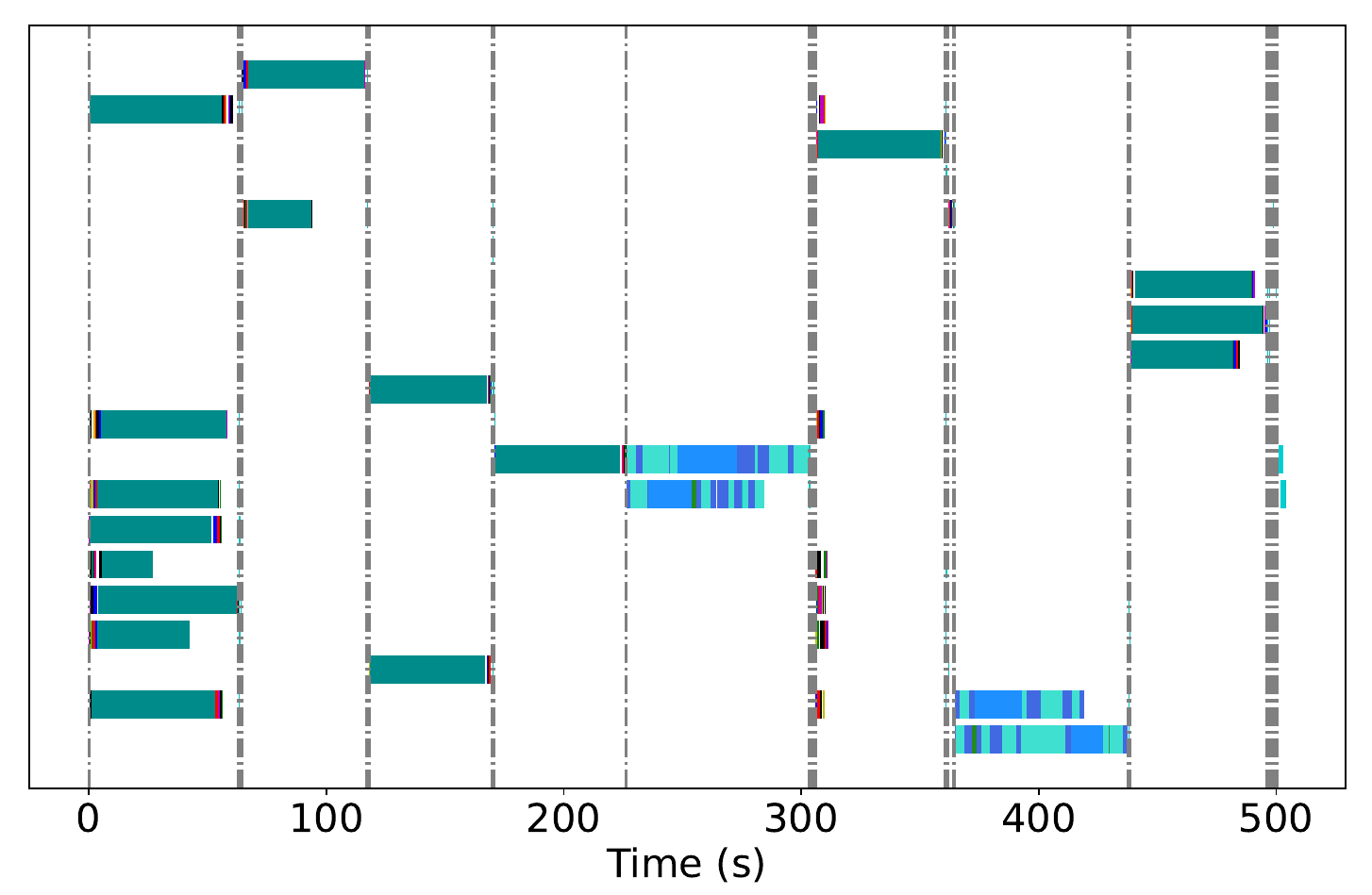}
    \caption{
    A snippet of the execution trace of a simulation. The x-axis shows the elapsed execution time, with each row representing an agent’s stream of LLM invocations. Colored bars denote different agent functions, and black dashed vertical lines indicate the completion of each step.    
    }
    \label{fig:bubble}
\end{figure}

\subsection{Motivation and Challenges}
\label{sub:motivation}

\paragraph{Imbalanced workload reduces available parallelism.}
Although line 5 in Algorithm~\ref{alg:procedure} suggests that parallelism can be achieved up to the number of agents, the effective parallelism is often much less primarily due to the imbalanced workload across agents. 
As illustrated in Figure~\ref{fig:bubble}, there are moments when many agents send out LLM requests uniformly. However, for the majority of the execution time, a few agents dominate each step, resulting in prolonged idle periods for many other agents who issue no LLM queries.
This sparsity is an inherent feature of simulation, as different agents have their own schedules and encountered events, making even distributions unlikely. 
Additionally, as shown in Figure~\ref{fig:bubble}, even for the same type of LLM calls, completions can vary significantly depending on inputs provided by different agents, further introducing imbalance.
Our measurements indicate that for a full-day simulation of $25$ agents, there are, on average, only $1.94$ concurrent LLM queries throughout the simulation.

\begin{figure}
    \centering
    \includegraphics[width=\linewidth]{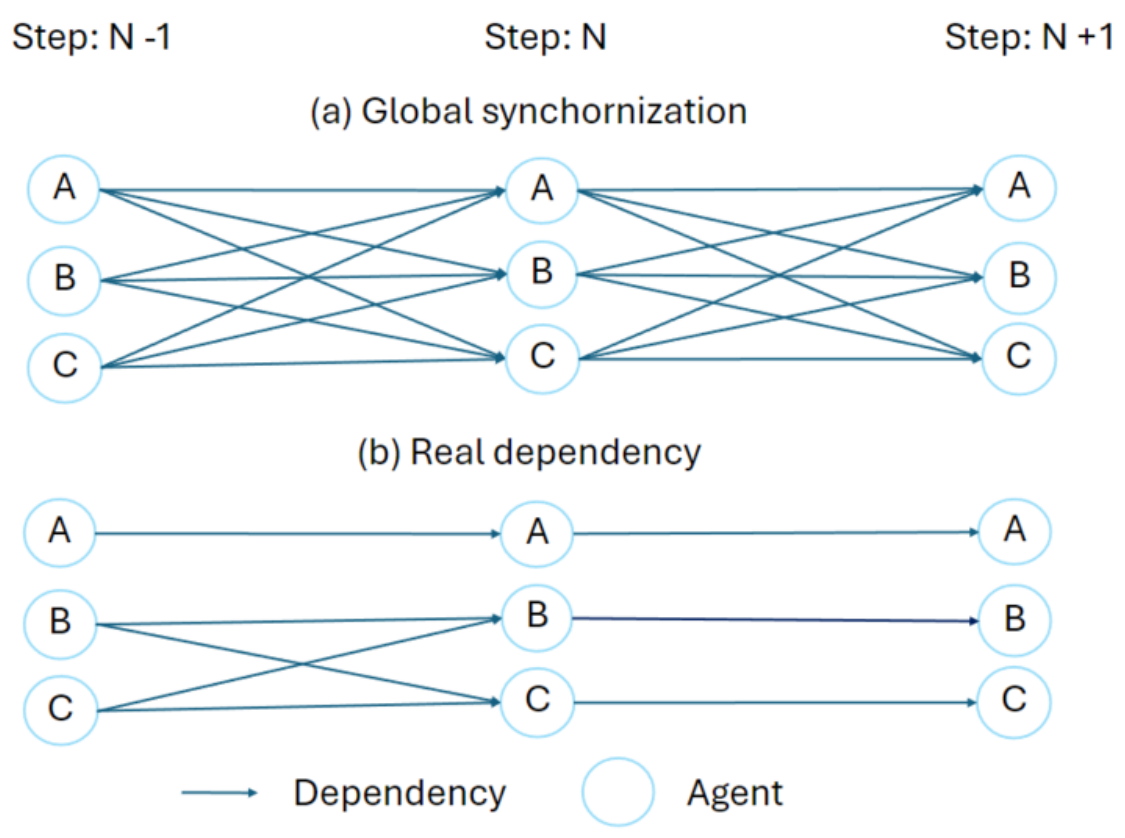}
    \caption{The dependency between agents' tasks is introduced by temporal causality. The top illustration shows an overly strict enforcement of this dependency, while the bottom illustration depicts a case of actual dependency.
    }
    \label{fig:causality}
\end{figure}

\paragraph{False Dependency.} 
While the workload imbalance across agents might be inevitable, low parallelism is not. 
We found that the primary cause of idleness is the overly strict enforcement of time causality. Requiring all events in one time step to complete before advancing to the next step introduces unnecessary dependencies.
As shown in Figure~\ref{fig:causality}, this approach creates an implicit all-to-all dependency across agents in consecutive steps. 
However, some agents, such as agent $A$, may be sufficiently isolated and unable to interact with others, thus not creating dependencies on agents like $B$ or $C$.
Our trace analysis for a whole day simulation of \GA 
indicates that, on average, each agent is dependent on only $1.85$ agents (including itself) from the prior step, far less than the default $25$.
The issue of false dependencies worsens as the agent count grows, as more false dependencies are enforced, diminishing the benefits of increased parallelism.
Although agents' behaviors are driven by responses from LLMs, which limits a scheduler’s ability to optimally manage dependencies without foresight, agent behavior is still somewhat predictable. Agents are constrained by their movement speed and limited action space, providing an opportunity to reduce most false dependencies through analysis of agents’ temporal-spatial relationships.

\paragraph{Requests of Different Priorities.} The dependency lens also reveals something unique about simulation compared to common LLM services like chatbots: there are long critical paths in the task of simulation, consisting of a chain of LLM calls that cannot be parallelized.
Therefore, requests have different priorities; those on the critical path should be served before non-critical requests to minimize the overall completion time as much as possible.

%% file: 4_design.tex
\section{Design of AI Metropolis}
\label{sec:design}

Motivated by observations described in \S\ref{sub:motivation}, we design \oursys,
an optimized simulation engine that serves as middleware between the developer-defined world and agents and the LLM serving engine, efficiently managing state updates and scheduling LLM queries.
By allowing agents to progress at varying speeds based on their LLM call loads, \oursys eliminates the need for frequent global synchronization, reducing false dependencies and maximizing parallelism.
Algorithm~\ref{alg:workflow} 
provides an overview of the new scheduling workflow adopted by \oursys, contrasting it with the traditional time step synchronized scheduling
shown in Algorithm~\ref{alg:procedure}.

\begin{algorithm}
\caption{AI Metropolis Scheduling Workflow}\label{alg:workflow}
\begin{algorithmic}[1]
    \STATE \textbf{Input:} target\_step, agents, world
    \STATE \textbf{Initialize:} base\_step $\gets$ 0, ready\_agents $\gets$ agents
    \STATE worker\_pool $\gets$ InitProcessPool(process\_routine)
    \STATE ready\_queue $\gets$ PriorityQueue()
    \STATE ack\_queue $\gets$ PriorityQueue()
    \STATE dependency\_graph $\gets$ Graph(agents)
    
    \vspace{0.3em} \hrule \vspace{0.3em}
    \COMMENT{\textbf{Controller}}
    \WHILE{base\_step $<$ target\_step} 
    \STATE ready\_clusters $\gets$ \textit{\textbf{geo\_clustering}}(ready\_agents)
    \STATE ready\_queue.put(ready\_clusters)
    \STATE ack\_cluster $\gets$ ack\_queue.get()
    \STATE ready\_agents $\gets$ update\_agents(agents, ack\_cluster)
    \STATE base\_step $\gets$ update\_base\_step(ack\_cluster)
    \ENDWHILE

    \vspace{0.3em} \hrule \vspace{0.3em}

    \COMMENT{\textbf{Worker}}
    \WHILE{base\_step $<$ target\_step} 
    \STATE cluster $\gets$ ready\_queue.get()
    \STATE actions $\gets$ [ ]
    \FORALL{agent \textbf{in} cluster}
        \STATE actions.append(agent.\textbf{\textit{proceed}}(world))
    \ENDFOR
    \STATE world.\textit{\textbf{resolve\_conflict\_and\_commit}}(actions)
    \STATE dependency\_graph.\textit{\textbf{update}}(cluster)
    \ENDWHILE

\end{algorithmic}
\end{algorithm}

\subsection{Overview}
Below we explain the essential terms in \oursys:
\begin{itemize}
    \item \textbf{Blocked}: An agent $A$ becomes \textit{blocked} if it has to wait for another agent $B$ to finish the current step before it can proceed, ensuring temporal causality and simulation correctness. This is formally defined in \S\ref{sub:dependency}. 
    \item \textbf{Coupled:} Agents $A$ and $B$, become \textit{coupled} if they are sufficiently close, interact with each other, and thus must proceed together. This is formally defined in \S\ref{sub:dependency}.

    \item \textbf{Cluster}: A cluster is a group of \textit{coupled} agents at the same step. Each agent can independently issue requests to LLMs within its \textit{proceed} function. However, the entire group needs to synchronize at the end of the step to resolve potential conflicts and avoid dependency violations as described in \S\ref{sub:dependency}.  

    \item \textbf{Worker}:  A worker is a process that handles one cluster, to proceed one step at a time. Within the worker process, each agent in the cluster operates in its own thread to communicate with the LLM serving engine and process its tasks. Workers are independent processes without synchronization between them, and the number of workers can be adjusted based on available CPU resources. 
    \item \textbf{Controller}: The controller is the main process of the simulation engine. After initializing the world, it periodically communicates with workers through two queues: it prepares tasks for workers via the \textit{ready\_queue} and confirms the completion of clusters from workers through the \textit{ack\_queue}.

\end{itemize}

During a simulation, workers continuously pull clusters from the \textit{ready\_queue}. After finishing the step for a cluster, a worker updates the dependency graph stored in a database for all agents in the cluster and then places the completed cluster in the \textit{ack\_queue} as a completion confirmation. 
Simultaneously, the controller continuously pulls notifications from the \textit{ack\_queue}. Each time it processes a confirmation, it queries the dependency graph to filter out the agents that are not \textit{blocked}, creating new ready clusters out of the ready agents and placing them into the \textit{ready\_queue} promptly. 
This workflow allows agents to advance steps ahead of others as long as dependency permits. Notably, both the \textit{ready\_queue} and \textit{ack\_queue} are priority queues that automatically sort tasks based on their associated steps.

We discuss the the dependency tracking mechanism in \S\ref{sub:dependency}, the spatiotemporal dependency graph realizing the mechanism in \S\ref{sub:graph}, agent clustering in \S\ref{sub:clustering}, priority scheduling in \S\ref{sub:priority}, and key implementation details in \S\ref{sub:implementation}.

\subsection{Dependency Tracking}
\label{sub:dependency}

As described in \S\ref{sub:motivation} and shown in Figure~\ref{fig:causality},
temporal causality introduces the dependence of tasks in a simulation. 
Using the language of computer systems, temporal causality creates an order of a set of reads and writes on a shared memory.
At the beginning of each step, agents read different parts of the environment and at the end, they commit writes to different parts. 
Therefore, the tasks of an agent across different steps must be serialized, as it reads what it wrote in the last step.
The dependency between tasks of different agents can then also be formulated as a read-after-write data dependency. For two agents $A$ and $B$, if $A$ is about to observe a part of the world at step $Step_A$, and $B$ is about to write to that part of the world at step $Step_B$. if $Step_B$ is smaller than $Step_A$, meaning the write is designed to happen before the read, then $A$ must wait for $B$ to complete step $Step_B$ before start the tasks in step $Step_A$.

In simulations, each agent typically perceives only a portion of the world, defined as the surrounding area within a specified size—denoted as $radius\_p$.
We also assume a maximum speed limit, denoted as $max\_vel$, governing both agents' movement and information propagation within each step.
For instance, in \GA, an agent perceives an area within a radius of $4$ grid units and can modify the status of an adjacent grid by interacting with an object or agent on it or moving to it.
This defines the region the agent can read from and write to. Consequently, to maintain read-after-write data dependency, the following condition must hold throughout the simulation:
\begin{align*}
&\forall \text{ agents } A, B, \text{ and their current steps } Step_A \text{ and } Step_B, \\
&\text{if } Step_A \neq Step_B, \text{then } dist(A, B) > radius\_p \\
& \hspace{1cm} + (|Step_A - Step_B| - 1) \times max\_vel
\end{align*}
where $dist(A, B)$ denotes the distance between $A$ and $B$.
This ensures that a read in a later step will never touch the region that a write in a prior step might influence. Intuitively, this means that agents never perceive other agents who exist at different times.
To enforce this condition, we can derive the following rules to determine the relationships between the tasks of agents.
The complete derivation can be found in Appendix~\ref{sec:appen} and the following are the rules \oursys uses: 
for any two agents $A$ and $B$ and their tasks in steps $Step_A$ and $Step_B$, 
\begin{itemize} 
    \item We define that $A$ and $B$ are \textit{coupled} if $Step_A = Step_B$ and $dist(A, B) \leq radius\_p + max\_vel$, which means they must be grouped into the same cluster and proceed to the next step together.
    \item We define that $A$ is \textit{blocked} by $B$ if $dist(A, B) \leq (Step_A - Step_B + 1) \times max\_vel + radius\_p$, which means $A$ cannot start its tasks in step $Step_A$ until $B$ has proceeded to the next step.
    \item A cluster can freely advance to the next step if none of its agents are blocked by any other agent.
\end{itemize}
Note that this set of rules are conservative, meaning the read in the later step will wait for all potential writes to a certain region to finish, even if the write does not occur eventually. While this may still preserve some false dependencies, we show in \S\ref{sec:eval} that it achieves performance close to the optimal. Additionally, this design does not require a data race detector and correction mechanism, making it more scalable and easier to implement.

\subsection{Spatiotemporal Dependency Graph}
\label{sub:graph}

\begin{figure}
    \centering
    \includegraphics[width=\linewidth]{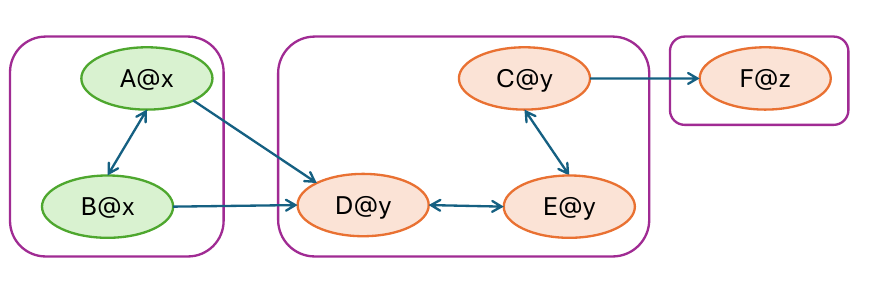}
    \caption{
    An example of a spatiotemporal dependency graph. Each node, such as A@x, represents an agent (A) at a specific time step (x). Single arrows indicate dependencies, while double arrows represent coupled relationships between agents. Purple boxes denote clusters of agents, where green nodes indicate agents that are ready for execution and orange nodes represent blocked agents.}
    \label{fig:spatiotemporal}
\end{figure}

The rules defined in \S\ref{sub:dependency} define a specialized dependency graph that tracks dependencies between agents.
Each node in the graph represents an agent, containing its temporal (time step) and spatial (coordinates on the map) information.
An edge $A \rightarrow B$ indicates that agent $B$ is currently \textit{blocked} by agent $A$, while $A \leftrightarrow B$ signifies that agents $A$ and $B$ are coupled as shown in Figure~\ref{fig:spatiotemporal}.

Our system maintains this graph in an in-memory database, and
whenever a worker advances a cluster to the next step, 
it applies the rules to re-examine the relationships of each agent in the cluster with any other relevant agents. Any changes are then written to the database. The examination and graph updates are encapsulated within a transaction to ensure data consistency. 
Afterward, the worker process places a completion confirmation into the \textit{ack\_queue}.

The dependency graph will be utilized by the controller process for efficiently identifying the agents that are not \textit{blocked}, allowing it to release maximum parallelism.

\subsection{Agent Clustering}
\label{sub:clustering}
When agents are close enough to each other, they perceive each other's actions committed in the last step. In other words, they collectively read what they wrote in the last step. 
They might also encounter write conflicts that must be resolved by developer-specified rules; for example, two agents might both want to use the bathroom, but only one can step in. These potential interactions couple them into a cluster that must proceed together.
Whenever there are new ready agents who are not blocked, the controller process runs \textit{geo\_clustering} to group coupled agents into clusters based on the rule described in \S\ref{sub:dependency}. 
If none of the members of a cluster are blocked, the cluster is considered ready and will be placed into the \textit{ready\_queue}.
Using clusters as the minimal synchronized units, as opposed to synchronizing all agents as described in  Algorithm~\ref{alg:procedure}, effectively reduces false dependencies and scheduling overhead.

\subsection{Priority Scheduling}
\label{sub:priority}
As motivated in \S\ref{sub:motivation}, 
allowing agents to process tasks associated with different time steps simultaneously creates requests of varying priorities. We found that the time step associated with a request serves as a good measure of its priority. 
A write operation in a prior step can block many reads in subsequent steps; intuitively, the smaller the time step, the more future actions it can potentially block.
To enhance parallelism, we maintain both the \textit{ready\_queue} and \textit{ack\_queue} as priority queues, prioritizing the execution of tasks from earlier steps.
No preemption during LLM inference is applied as that might cause extra overhead~\cite{sheng2023fairness}.
We demonstrate the effectiveness of this priority scheduling in \S\ref{sub:breakdown}. 

\begin{figure*}[t!]
	\centering
	\begin{subfigure}{0.32\textwidth}
		\centering
		\includegraphics[width=0.95\textwidth]{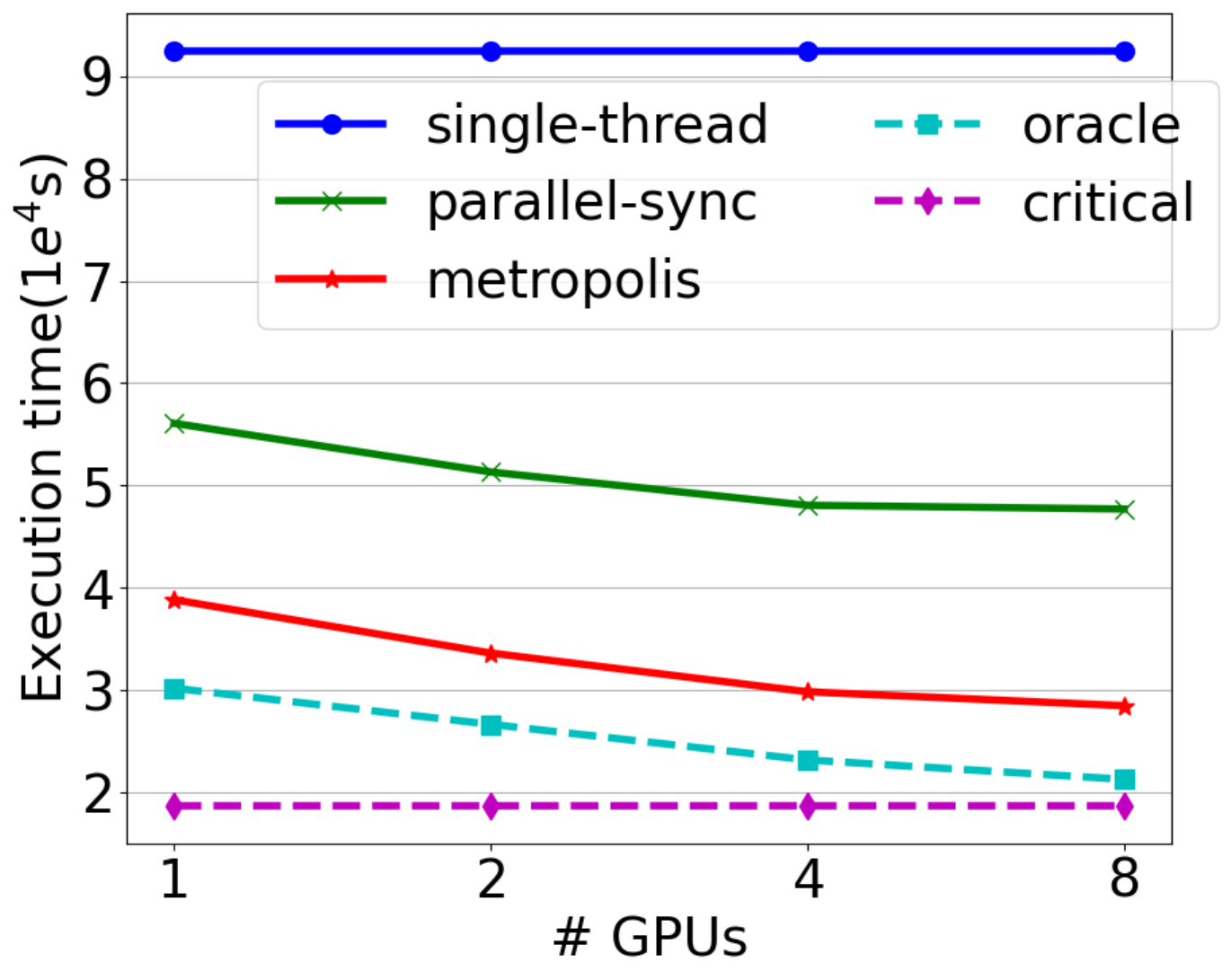}
		\caption{\small Simulation using Llama-3-8b-instruct \\ on NVIDIA L4 GPU}
		\label{fig:l4_exe}
	\end{subfigure}
    \begin{subfigure}{0.32\textwidth}
		\centering
		\includegraphics[width=0.95\textwidth]{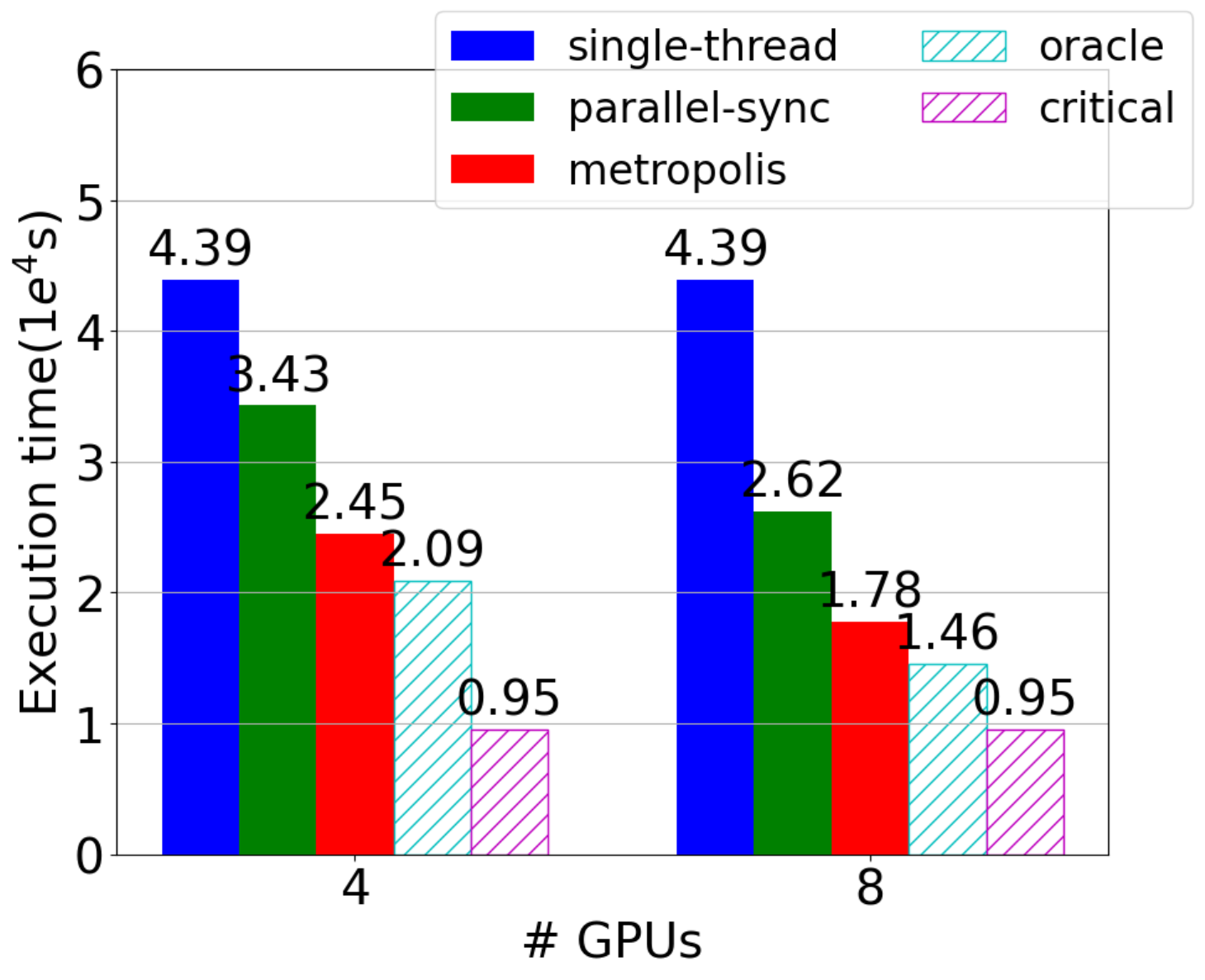}
		\caption{\small Simulation using Llama-3-70b-instruct on NVIDIA A100 GPUs}
		\label{fig:a100_exe}
	\end{subfigure}
    \begin{subfigure}{0.32\textwidth}
		\centering
		\includegraphics[width=\textwidth]{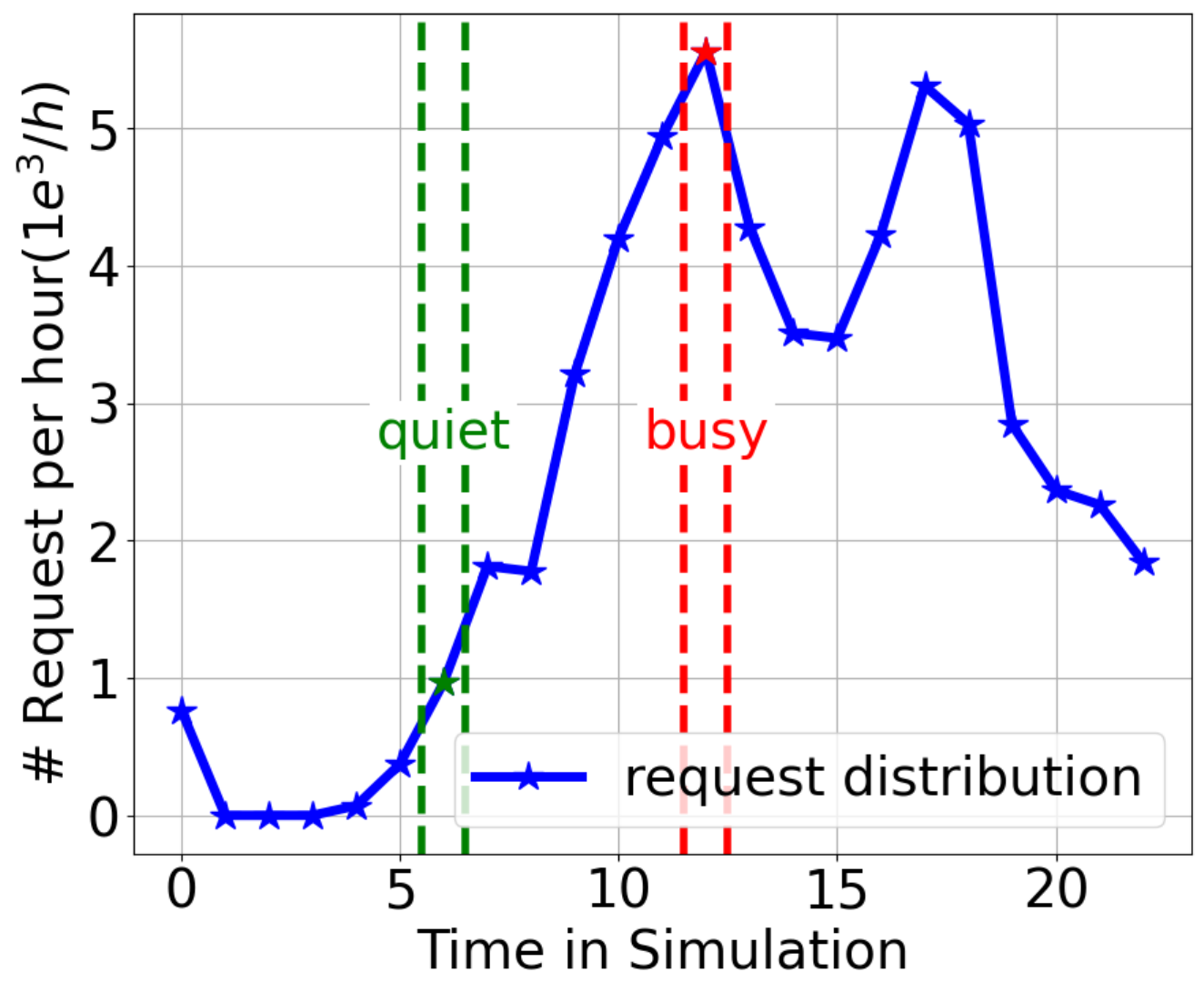}
		\caption{\small LLM query distribution in simulation}
		\label{fig:call_distribution}
	\end{subfigure}
	\caption{\small (\ref{fig:l4_exe}, \ref{fig:a100_exe}) End-to-end 25 agents full day simulation completion time with different number of GPUs.
    (\ref{fig:call_distribution}) shows the distribution of LLM calls over the simulated hours, note the low activity period during 1am-4am is because all agents are sleeping. 
    }
	\label{fig:25agent}
\end{figure*}

\subsection{System Implementation}
\label{sub:implementation}
In addition to the design choices around dependency management, clustering, and priority scheduling that enable \oursys to expose more parallelism from the simulation, it is also worth highlighting some of the design choices that make \oursys scalable:

\paragraph{Proper Mapping of Parallelism.} Choosing the right programming abstraction for different tasks is critical to scalability. \oursys employs threads for agents, as the need for synchronization within clusters requires low-overhead communication. Processes are used for the controller and workers to avoid the Python GIL and to enable scaling beyond a single machine.

\paragraph{Light and Fast Critical Path on the Controller Process.} All tasks on the critical path are implemented in C++ to reduce overhead and bypass the limitations of the Python GIL. Furthermore, heavy lifting, including complex agent processing logic and dependency graph updates, is offloaded to concurrent workers. This approach lightens the critical path for the controller process, minimizing workers' waiting time for task allocation.

\paragraph{Scalable I/O.} Except for \textit{ready\_queue} and \textit{ack\_queue}, all inter-process synchronization are handled through an in-memory~\cite{redis} database.
This database also handles transactional updates for all simulation states and stores instrumentation data, supporting automatic scalability beyond a single node.

\paragraph{Decoupling Simulation Engine from LLM Serving Engine.} In \oursys, only workers communicate with the LLM serving engine through a thin shim layer, providing easy observability and scalability. 

We implemented the core of \oursys in about 1k lines of C++ and 2k lines of Python code. An additional 3k lines of Python code (excluding assets and prompts) were written to port the \GA simulation using our interfaces. This is about 
$50\%$ of the code compared to the original implementation, achieving up to an order of magnitude speed-up and promising scalability, as demonstrated in \S\ref{sec:eval}.

%% file: 5_evaluation.tex
\section{Evaluation}
\label{sec:eval}

In the evaluation, we aim to answer the following questions:
\begin{itemize}
    \item Does \oursys effectively enhance parallelism by tracking real dependencies, and how does this translate to shorter completion times?
    \item Does \oursys scale as the size of the simulated world increases and the number of agents grows?
    \item Given that \oursys does not eliminate all false dependencies as illustrated in \S\ref{sub:dependency}, how well does it perform compared to the optimal solution?
\end{itemize}

We describe the experimental setup in \S\ref{sub:setup} and discuss the performance results of full-day simulations at a small scale in \S\ref{sub:original}, which uses the same simulation settings reported in the \GA paper. We then examine the performance comparisons as the size of the world increases and the number of agents grows to a thousand, assessing scalability in \S\ref{sub:scaling}. Finally, we conduct a performance breakdown in \S\ref{sub:breakdown} to demonstrate the effectiveness of priority scheduling.

\subsection{Methodology}
\label{sub:setup}
\paragraph{Serving Engine.} We use SGLang~\cite{zheng2023efficiently} (v0.1.17) as the LLM serving engine, as it is not only one of the state-of-the-art LLM serving engines but also lightweight and easy to instrument and modify.
For consistent and stable performance benchmark results, we turned off its automatic common prefix caching feature; however, enabling the cache generally provides about a $20\%$ throughput gain across all settings.

\paragraph{Model and Hardware Platform.} 
We benchmarked \oursys with various models and GPUs to assess its effectiveness across different sizes and complexities.
We chose state-of-the-art open-source LLMs from the Meta Llama-3 instruct series~\cite{llama3}. Community benchmarks~\cite{chiang2024chatbotarenaopenplatform} indicate that the smallest 8B model already surpasses the ChatGPT-3.5 model used in the original \GA paper, making it ideal for performance evaluation.
We benchmarked our system with both the 8B and 70B models. The 8B model offers a lightweight deployment option, while the 70B model provides advanced capabilities, though at a higher cost. For the Llama-3 8B experiments, we used NVIDIA L4 GPUs on GCP G2 instances, scaling from one to eight GPUs to assess data parallelism. For the Llama-3 70B experiments, we used NVIDIA A100-80GB GPUs, applying tensor parallelism across four GPUs, and expanding to eight GPUs for a hybrid data and tensor parallelism configuration.
Additionally, we benchmarked \oursys using the Mixtral-$8\times7$B-Instruct-v0.1~\cite{mixtral} model, a mixture of expert models, on the same A100 platform which can leverage higher data parallelism to reveal more performance characteristics.
\paragraph{Traces.} 
We collected workload traces for 40 simulation days of \GA  by instrumenting the original implementation~\cite{generativecode} and running it multiple times using the same settings reported in the paper. OpenAI GPT-3.5 API service~\cite{gpt35} was used as the LLM engine as the same setting in the paper.
On average, each simulation day's trace consists of $56.7k$ LLM call events. Each event includes the input prompt, configurations, LLM response, calling step, and caller's identity. A separate trace file tracks the agent's movements throughout the simulation. The average length of input tokens is $642.6$, and the average length of output tokens is $21.9$.
We conducted the performance benchmark using the replay mode of \oursys, faithfully replaying these traces to ensure the same movements, interaction patterns, inputs, and the same length of generation output by setting \textit{ignore\_eos} in SGLang for comparable and stable performance results.

\begin{figure*}[htb!]
    \centering
    \begin{subfigure}{\textwidth}
        \centering \includegraphics[width=0.95\textwidth]{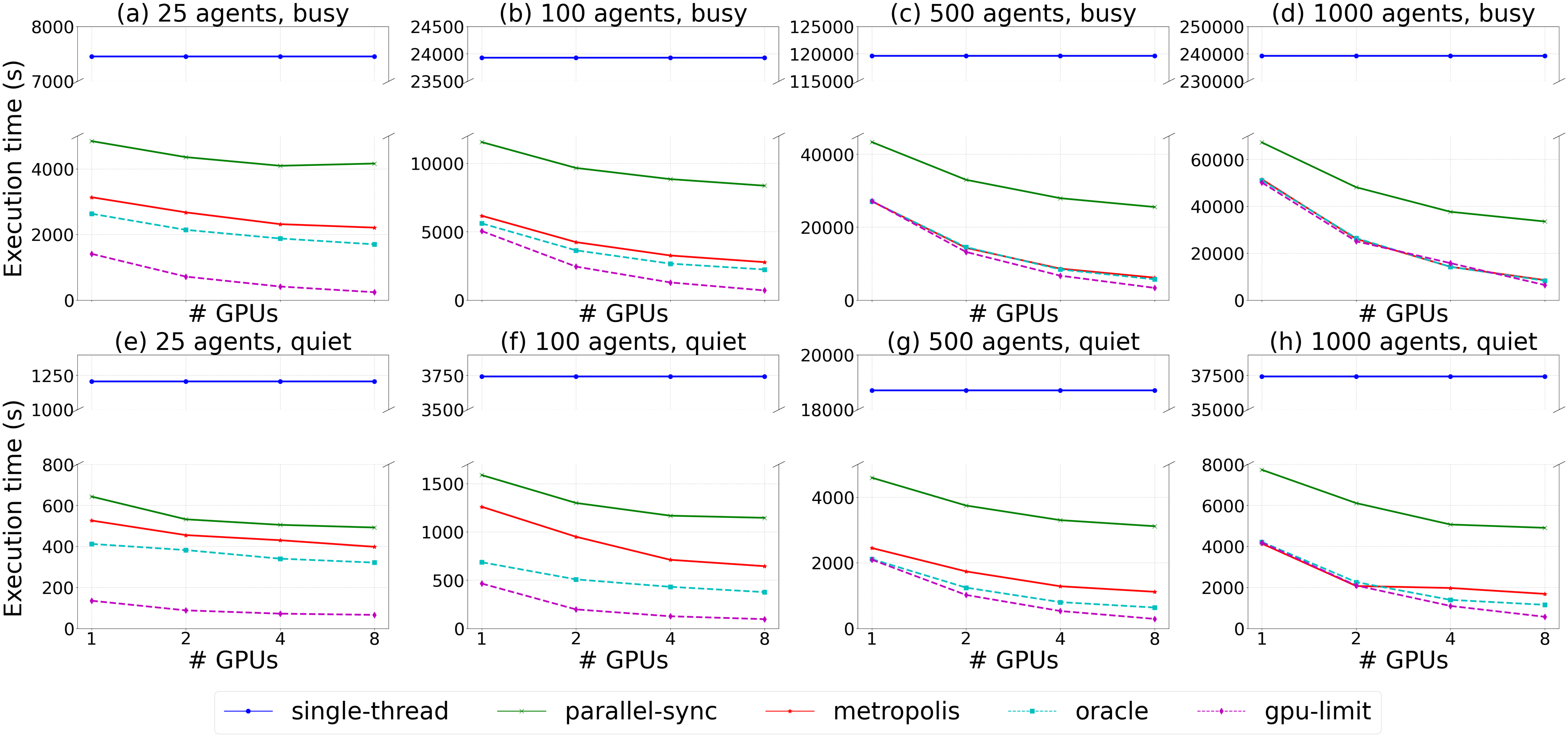}
    \end{subfigure}
    \caption{
    Benchmark of busy (12 a.m. - 1 p.m.) and quiet (6 a.m. - 7 a.m.) hours using Llama-3-8b-instruct on NVIDIA L4 GPUs, with agent counts scaled from 25 to 1000. Single-thread results for 500 and 1000 agents are projected based on workload estimations.
    }
    \label{fig:l4_scaling}
\end{figure*}

\subsection{Full Day Simulation in SmallVille}
\label{sub:original}

We benchmark \oursys using the same setup described in the \GA paper, which involves 25 agents within a world named SmallVille, a 100x140 grid, running for a full simulation day.

The following experiment settings are used in benchmark:
\begin{itemize}
    \item \textit{single-thread} employs a single thread to handle simulation states and issue LLM requests, as per the design adopted by the original implementation to simplify simulator implementation. No parallelism is exposed for LLM requests from different agents.
    \item \textit{parallel-sync} is a stronger baseline where all agents operate within the same time step and issue LLM requests independently, though global synchronization limits achievable parallelism as discussed in \S~\ref{sub:motivation}. We implemented this baseline as a mode of \oursys.
    \item \textit{oracle} represents the optimal dependency management solution for comparison. This setting constructs an optimal dependency graph by analyzing the full trace and mining all necessary dependencies based on agent interactions. For example, if two agents appear in each other's observation space, they synchronize before and after the step to ensure temporal causality. This setting is unattainable in real systems and serves to illustrate the potential improvement of dependency management. By having an optimal dependency graph, the most available parallelism will be released.
    \item \textit{critical}refers to the critical path of the simulation, extracted from the optimal dependencies used in the oracle setting. It identifies the path containing the most LLM input and output tokens, setting an lower bound of completion time regardless of available resources.

\end{itemize}


First, \oursys outperforms the \textit{single-thread} and \textit{parallel-sync} baselines by $2.38\times$ and $1.44\times$ on a single L4 GPU. 
As more GPUs are added, requiring greater parallelism, the speedup increases to $3.25\times$ and $1.67\times$ respectively on 8 GPUs. 
We also measured the achieved parallelism for each simulation by averaging the number of outstanding requests over the execution time, where \oursys reached $3.46$, compared to $0.95$ for \textit{single-thread} and $1.94$ for \textit{parallel-sync} on 8 GPUs.
These results align with the observed speedups, as greater parallelism improves GPU utilization and overall performance.

\oursys also approaches \textit{oracle} performance, reaching $74.7\%$ of the oracle performance on 8 GPUs and up to $82.9\%$ on a single GPU.
The gap stems from the longer critical path in \oursys compared to \textit{oracle}, as it conservatively restricts certain agents from advancing to prevent potential temporal causality violations, as elaborated in \S\ref{sec:design}. We further discuss this gap in \S\ref{sec:discussion}.

A similar trend is observed in benchmarks conducted on A100 GPUs with larger models. \oursys achieves a $2.45\times$ and $1.45\times$ speedup compared to \textit{single-thread} and \textit{parallel-sync}, respectively, and attains $82\%$ of the \textit{oracle} performance on 8 GPUs.
Additional speedups are anticipated with higher data parallelism, given the \textit{oracle-to-critical} ratio of $64.7\%$ on A100s versus $88\%$ on L4 GPUs, as memory demands for 70B models ($8.75\times$ higher) limit processing capacity.

\subsection{\textbf{Scaling} up to a Thousand Agents}
\label{sub:scaling}
\begin{figure*}[htb!]
    \centering
    \begin{subfigure}{\textwidth}
        \centering
        \includegraphics[width=0.95\textwidth]{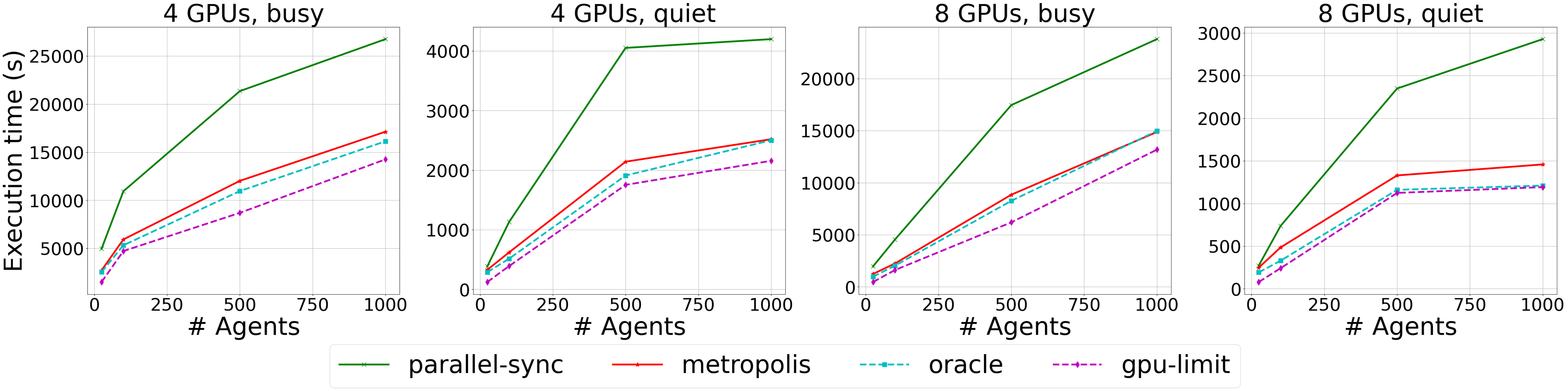}

        \label{fig:KV_Cache_KIVI}
    \end{subfigure}
    \caption{
    Benchmark of busy (12a.m. - 1p.m.) and quiet (6a.m. - 7a.m.) hour using Llama-3-70b-instruct on NVIDIA A100 GPUs with scaling number of agents rom 25 to 1000.}
    \label{fig:a100_scaling}
\end{figure*}

\begin{figure}[ht]
    \centering
    \includegraphics[width=0.95\linewidth]{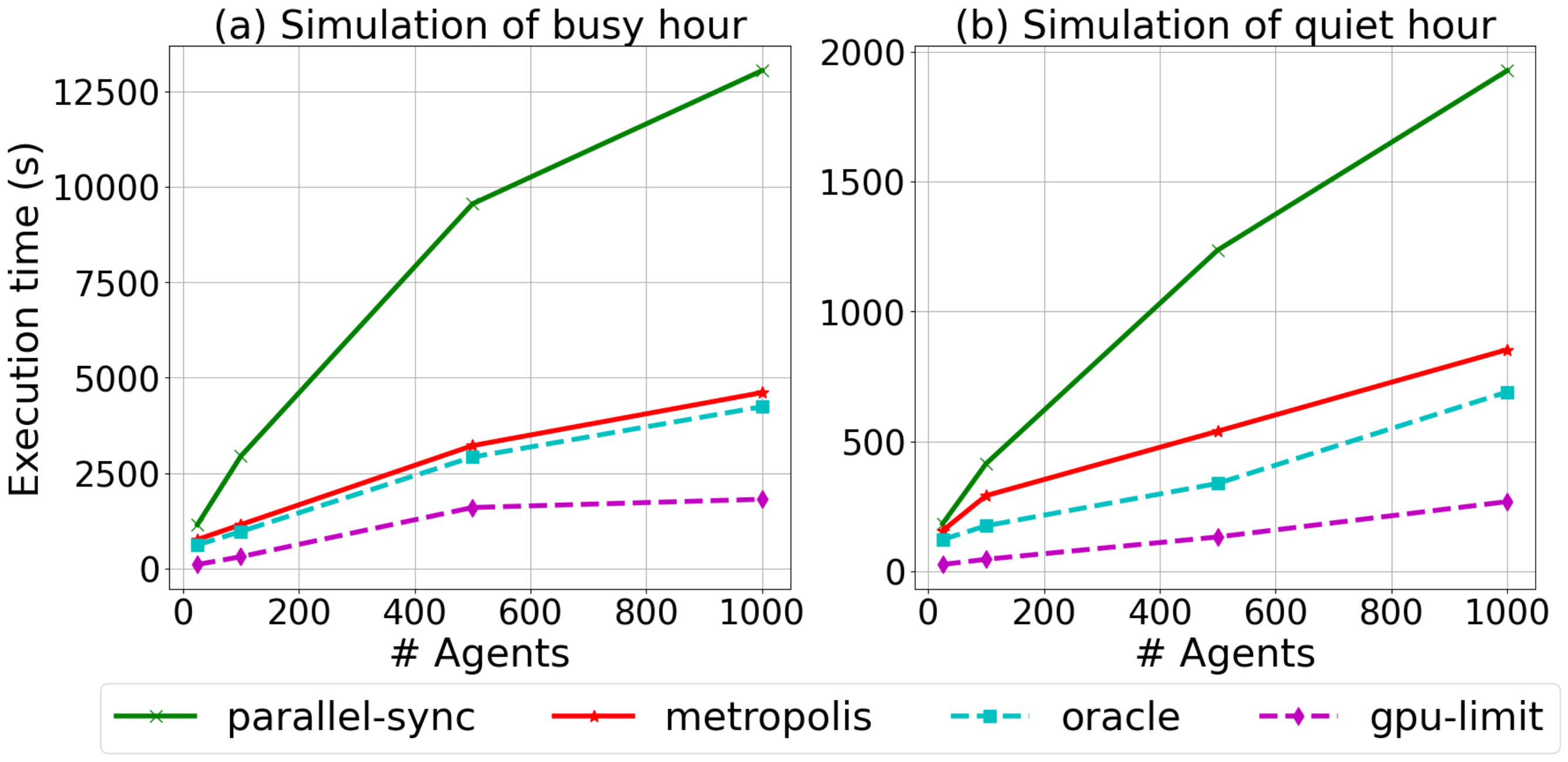}
    \caption{Benchmark of Mistral 8$\times$7 on 8 A100 GPUs, with agent counts scaled from 25 to 1000. 
    }
    \label{fig:moe}
\end{figure}

Given the limited research on accommodating hundreds of agents, we simulate a larger environment by concatenating multiple SmallVilles into a single, large ville for benchmarking. 
Agents in each segment replay different traces that we collected independently, but they operate within the same time and space.
Since the concatenation approach introduces straightforward parallelism, rather than focusing on the critical path, which is artificially shortened due to the lack of interaction between different parts of the large ville, we introduce \textit{no-dependency} as a more suitable lower bound for completion time when scaling agents. In this setting, all LLM calls can be issued simultaneously, maximizing hardware utilization. In Figure~\ref{fig:l4_scaling},~\ref{fig:moe} and \ref{fig:a100_scaling}, the \textit{gpu-limit} uses the shorter completion time of the \textit{critical} and \textit{no-dependency} settings.
Moreover, for benchmark with a larger number of agents, we opted to focus on two specific intervals from an entire day's simulation, as illustrated in Figure~\ref{fig:call_distribution}:
the busy hour (12 PM - 1 PM, approximately 5,000 calls) and the quiet hour (6 AM - 7 AM, approximately 800 LLM calls). 
This setup shortens experiment time and highlights scaling effects across different workloads, where busy hours feature long conversations, and quiet hours are mainly routine activities with less LLM queries as agents just wake up.

The benefits of \oursys increase with increasing numbers of agents.
Figure~\ref{fig:l4_scaling} shows that \oursys achieves closer performance to \textit{oracle} as the number of agents increase: it achieves $90\%$ of \textit{oracle} on one GPU with 100 agents, reaching parity with \textit{oracle} at 500 agents. On 8 GPUs, \oursys improves from $53.1\%$ to $97.0\%$ of \textit{oracle} across settings.
Speedups over \textit{single-thread} and \textit{parallel-sync} also scale with agent count, increasing from $3.37\times$ and $1.88\times$ at 25 agents to $19.5\times$ and $4.15\times$ at 500 agents. Unlike \textit{single-thread}, which cannot leverage parallelism, and \textit{parallel-sync}, which suffers from costly synchronization, \oursys utilizes parallelism more effectively, maximizing speedup as agent count grows.

After reaching peak speedup over \textit{parallel-sync} at 500 agents, the speedup plateaus, slightly decreasing to $3.94\times$ at 1000 agents. This is because, as agent count grows relative to available computational resources, even less efficient dependency management achieves adequate hardware utilization. Meanwhile, \oursys reaches $97\%$ of \textit{oracle} performance, indicating that additional parallelism is less effective.
This trend appears earlier on a single L4 GPU, where computational resources are more limited. \oursys achieves a maximum speedup of $1.87\times$ over \textit{parallel-sync} at 100 agents, tapering to $1.60\times$ as \oursys's performance approaches \textit{oracle}—from $90.9\%$ at 100 agents to $100\%$ at 500 agents.

Similar trends appear in the quiet hour benchmark, as shown in Figure~\ref{fig:l4_scaling}, with some variation: the lighter and less frequent LLM calls in the quiet hour benchmark reduce the synchronization overhead for \textit{parallel-sync}, allowing more parallelism. 
As a result, \oursys shows a smaller speedup over \textit{parallel-sync} with the same agents and GPUs—for instance, $1.28\times$ in the 25-agent, 8-GPU setting, where achieved parallelism is $2.25$ for \textit{parallel-sync} and $2.80$ for \oursys. By comparison, the busy hour benchmark achieves parallelism values of $1.89$ and $3.74$ on the same setting, respectively. Nevertheless, as the number of agents increases, the speedup for \oursys rises from $1.28\times$ to $2.79\times$ at 500 agents on 8 GPUs.

Similar trends also hold for larger models on 8 A100 GPUs. 
\oursys peaks at a $1.97\times$ speedup over \textit{parallel-sync} with 500 agents in the busy hour benchmark and $2.01\times$ in the 1000-agent quiet hour benchmark, as shown in Figure~\ref{fig:a100_scaling}.
To further explore model variability, we benchmarked the Mistral MoE $8\times7b$ model on the same 8 A100 platform, which uses $80\%$ of a 70b model’s memory with lighter I/O and computation. With the $8\times7b$ MoE model, we observe higher peak speedups of $2.97\times$ and $2.29\times$ over \textit{parallel-sync} at 500 agents for busy and quiet hour benchmarks, respectively, due to greater resource availability on the GPUs, which allows for better parallelism utilization.

\subsection{Priority Scheduling Breakdown}
\label{sub:breakdown}

\begin{table}[h!]
    \centering
    \begin{tabular}{lcccc}
        \toprule
        \# GPUs & \multicolumn{2}{c}{\text{metropolis}} & \multicolumn{2}{c}{\text{oracle}} \\
        \cmidrule(r){2-3} \cmidrule(r){4-5}
        & 4 & 8 & 4 & 8 \\
        \midrule
        w/ priority (s) & 8611 & 6148 & 8392 & 5683 \\
        w/o priority (s) & 8942 & 7114 & 8484 & 5689 \\
        Speedup (\%) & 3.84\% & 15.7\% & 1.10\% & 0.11\% \\
        \bottomrule
    \end{tabular}
    \caption{Performance breakdown of \textit{metropolis} and \textit{oracle} with and without priority scheduling on L4 GPUs. The first two rows are completion time in seconds.}
    \label{tab:priority}
\end{table}

All the experiments discussed so far have priority scheduling enabled, where every request includes a step count, and requests with smaller counts have higher execution priority. This applies to the \textit{oracle} baseline as well.
We repeated the experiment of busy hours with 500 agents on 4 and 8 L4 GPUs for \oursys and the \textit{oracle}, but with priority scheduling turned off. 

As shown in Table~\ref{tab:priority}, priority scheduling does not significantly impact performance of \textit{oracle} because it already achieves sufficient parallelism, and its dependency graph is sparse as discussed in \S\ref{sub:motivation}, making priority less critical for unlocking additional parallelism.
In contrast, we observed up to a 15.7\% speedup for \oursys with priority scheduling. 
This is because the conservative rules defined in \S\ref{sub:dependency} make agents falling behind to block others more frequently. 
Priority scheduling reduces this blocking, allowing \oursys to perform closer to the \textit{oracle}. 
With priority enabled, the average achieved parallelism in the 500-agent, 8-GPU benchmark increases from 41.9 to 50.9 for \oursys, compared to a minor increase from 69.4 to 69.9 for \textit{oracle}.

%% file: 6_related.tex
\section{Related Work}

\paragraph{LLM Agent Society Simulation and Multi-agents Collaboration.}
With increasing interest in LLM agents, several recent frameworks, such as Camel~\cite{li2023camel}, AutoGPT~\cite{AutoGPT}, and OpenAI Swarm~\cite{openai_swarm_2024}, have emerged to simplify the development of multi-LLM agent interactions. However, these frameworks primarily provide interfaces for connecting LLM agents but lack a shared environment or state synchronization, making multi-agent interactions more akin to microservices connected via remote procedure calls.
In contrast, \GA and similar explorations~\cite{wang2023voyager, gong2023mindagent, altera_ai_npcs} represent a different approach, which we call \textit{AI society}. In these systems, agents exist within a virtual world, interacting with both each other and the environment. 
However, this line of research typically emphasizes agent architectures and prompt engineering, often resulting in a lock-step simulation process for simplicity, which can lead to slower simulations.
AI Town~\cite{aitown}, although inspired by generative agents and resembling AI society frameworks, offers an open-source platform where the environment is only minimally interactive for agents. Agent interactions are limited to simple rule-based operations, which constrains its versatility.
\oursys tackles the performance challenges inherent to AI society simulation, offering a solution that enhances efficiency without compromising the functionality.

\paragraph{General Multi-agent Simulation.}
While LLM agent simulation may appear superficially similar to general multi-agent simulations, such as those used in reinforcement learning~\cite{gym, zhu2024pearl, shacklett2023extensible} and multi-agent processing~\cite{emau2011multi}, they present fundamentally different scheduling challenges due to the high per-agent computational demands and significant workload imbalances inherent in LLM execution, as discussed in \S\ref{sub:motivation}.
These unique demands necessitate specialized scheduling strategies.
\oursys is designed to offer a user experience comparable to that provided by established reinforcement learning frameworks, while seamlessly managing the additional complexities behind the scenes.

\paragraph{LLM Serving Optimizations.}
An active line of research~\cite{yu2022orca, kwon2023efficient, agrawal2023sarathi, zheng2023efficiently, sheng2023s, chen2024punica} in LLM inference optimization has been making continuous advancements in increasing throughput and reducing latency from various angles. These engines typically achieve optimal performance when there are ample requests available for scheduling but generally do not address request dependencies.
\oursys complements these efficient engines by increasing parallelism within the simulation, thereby enhancing overall throughput. Since \oursys decouples the simulation engine from the serving engine, improvements in serving engines directly boost simulation throughput.
While a substantial body of research exists on general inference and serving, it remains largely orthogonal to our work. 
Some studies address dependencies among LLM requests, such as \cite{kim2023llm} for function calling and \cite{zheng2023efficiently} for LLM programs to achieve higher parallelization. However, these approaches assume rigid dependencies, unlike \oursys, which focuses on reducing false dependencies to further enhance efficiency.

%% file: 7_future.tex
\section{Discussion and Future Work}
\label{sec:discussion}

\paragraph{Conservative or Speculative Execution.} 
As discussed in \S\ref{sub:dependency}, \oursys\ applies conservative rules to prevent causality violation, which can extend the critical path and reduce overall parallelism.
Nonetheless, as demonstrated in \S\ref{sec:eval}, \oursys\ achieves performance close to the \textit{oracle} setting in most cases, primarily due to effective priority scheduling. Although scenarios exist where additional parallelism could enhance performance, the performance gap between \oursys\ and the \textit{oracle} remains minor, as evidenced by the quiet hour benchmark in \S\ref{sub:scaling}. 
This gap, however, highlights opportunities to further improve \oursys. Introducing speculative execution with race detection could potentially bridge this gap, although doing so may challenge the system's scalability principles—a direction we leave open for future work.

\paragraph{Offline and Interactive.}
While \oursys currently focuses on throughput for offline simulation, its core principles including fine-grained dependency management and priority scheduling are also applicable to interactive environment like video games. 
The key difference between a real game and a simulation lies in interactivity, which imposes latency requirements.
We view games like The Sims~\cite{thesims} as hybrids of interactive and offline simulation components. 
The part with which the player interacts must have low latency for real-time responsiveness, while background agents should operate in a simulation manner to exhibit realistic social behaviors.
An important future work of \oursys is to support such hybrid deployment, balancing request prioritization to reduce latency in interactive components and optimize throughput in background simulations.

\paragraph{Applications of \oursys.}
Although this paper highlights \GA as the primary use case, \oursys offers broad applicability across various domains:
1) As a pioneering framework for simulating social behaviors with LLMs, \GA has influenced many studies, expanding its impact. An efficient engine for \GA reaches a wide audience, as the observed imbalanced load and sparse dependency patterns are common across human-like simulations. Additionally, the dependency rules in \oursys can be adapted to other environments by adjusting parameters like perception radius and movement speed to suit different dynamics.
2) While our derivations focus on temporal-spatial relationships in Euclidean space, they can extend to non-Euclidean spaces, such as social networks, highlighting \oursys's flexibility for diverse simulations needing dependency management.

\paragraph{Online APIs and Local Models.} While proprietary APIs like GPT-4~\cite{openai} and Claude 3~\cite{anthropic} lead in performance, open-source models are on the rise. \oursys supports local model serving with optimized dependency management for faster, cost-effective simulations, yet remains compatible with online APIs, enhancing parallelism and simplifying state management for users.

\section{Acknowledgment}
We thank Joon Sung Park, Brennan Shacklett, Mark Zhao, Athinagoras Skiadopoulos, Qizheng Zhang, Piero Molino and \cite{altera2024} for their valuable discussions and feedback.
This research was partly supported by the Stanford Platform Lab and its affiliates, and by ACE, one of the seven centers in JUMP 2.0, a Semiconductor Research Corporation (SRC) program sponsored by DARPA. This research was also partly supported by NSF CNS-2047283.

%% file: appendix.tex
\section{Rules Derivation}
\label{sec:appen}
 
At any point during the out-of-order simulation, the state includes each agent's location and the step they have executed. For a given state, let $dist(A, B)$ represent the distance between agents $A$ and $B$, 
$radius\_p$ denote the radius of an agent's perception, and $max\_vel$ denote the maximum speed of agent movement and information propagation per step.
According to Section~\ref{sub:dependency}, a valid execution needs to make sure the following condition holds at any state:
\begin{align*}
&\forall \text{ agents } A, B, \text{ and their current steps } Step_A \text{ and } Step_B, \\
&\text{if } Step_A \neq Step_B, \text{then } dist(A, B) > radius\_p \\
& \hspace{1cm} + (|Step_A - Step_B| - 1) \times max\_vel
\end{align*}

To satisfy the condition, we derive the following simulation conditions to ensure the state remains valid.
Notice that our simulation conditions are over-estimations, which is sound in correctness but not necessarily complete.

Starting from any valid state, given any two agents $A$ and $B$ at steps $Step_A$ and $Step_B$, respectively, we derive the simulation conditions of $A$ by case study.
Let $A'$ denote the new state of $A$ after one more step so that $dist(A', B)$ denotes the new distance between $A$ and $B$ after the next step of $A$.

\begin{itemize}
    \item Assume steps $Step_A = Step_B$. 
    $A$ is allowed to proceed to the next step if a valid state is reached after one further step of $A$. Formally, there should be
\begin{align*}
    dist(A', B) > &(Step_A-Step_B + 1 - 1) \times max\_vel \\
    & + radius\_p
\end{align*}    
    Since $dist(A', B) \geq dist(A, B) - max\_vel$, we need:
    \[dist(A, B) - max\_vel > radius\_p\]
    Therefore, on the other side, $A$ and $B$ must stay at the same step if:
    \[dist(A, B) \leq max\_vel + radius\_p\] 
    which means they are \textit{coupled} and can either wait together or proceed together.
    
    \item Assume steps $Step_A > Step_B$.
    There should be
\begin{align*}
    dist(A', B) > &(Step_A-Step_B + 1 - 1) \times max\_vel \\
    & + radius\_p
\end{align*}    
    Since $dist(A', B) \geq dist(A, B) - max\_vel$, we need:
\begin{align*}
    dist&(A, B) - max\_vel \\&>  radius\_p 
    + (Step_A-Step_B) \times max\_vel
\end{align*}
    Therefore, on the other side, $A$ got \textit{blocked} by $B$ if
\begin{align*}
    dist(A, B) \leq & (Step_A-Step_B+1) \times max\_vel
    \\ &+ radius\_p
\end{align*}    
    \item Assume steps $Step_A < Step_B$.
    There should be
\begin{align*}
dist(A', B) > &(Step_B-Step_A-1- 1) \times max\_vel
\\ &+ radius\_p
\end{align*}    
    Since $dist(A', B) \geq dist(A, B) - max\_vel$, we need:    
\begin{align*}
di&st(A, B) - max\_vel \\
&> (Step_B-Step_A-2) \times max\_vel + radius\_p
\end{align*}       
\begin{align*}
\Rightarrow dist(A, B) > &(Step_B-Step_A-1) \times max\_vel
\\ &+ radius\_p
\end{align*}
    which is the same valid condition of the current state.
    Therefore, A is not \textit{blocked} by any future agents.
\end{itemize}